\documentclass[]{tJOT2e}

\renewcommand{\eqref}[1]{~(\ref{#1})}
\newcommand{\upd}{\mathrm{\,d}}
\usepackage{amsmath}
\usepackage{amssymb}
\usepackage{wasysym}
\usepackage{xcolor}
\usepackage{lineno}
\begin{document}
\doi{10.1080/14685248.YYYYxxxxxx}
 \issn{1468-5248}
 \jvol{00} \jnum{00} \jyear{2014}

\markboth{Yongxiang HUANG}{Journal of Turbulence}
\linenumbers

\title{Detrended Structure-Function in Fully Developed Turbulence}

\author{Yongxiang HUANG$^{\ast}$\thanks{$^\ast$Corresponding author. Email: yongxianghuang@gmail.com
\vspace{6pt}} \\\vspace{6pt}   {\em{Shanghai Institute of Applied Mathematics and Mechanics, Shanghai Key Laboratory of Mechanics in Energy Engineering, Shanghai University,
Shanghai 200072, People\rq{}s Republic of China}}\\\vspace{6pt}\received{v3.2 released February 20XX} }

\maketitle

\begin{abstract}
The  classical  structure-function (SF) method in fully developed turbulence or for 
scaling processes in general is  influenced by large-scale energetic 
structures, known as infrared effect. Therefore, the extracted scaling exponents 
$\zeta(n)$ might be biased due to this  effect. 
In this paper, a  detrended structure-function (DSF)  method is proposed to 
extract scaling exponents by constraining the influence of large-scale structures. This is accomplished by removing a $1$st-order 
polynomial fitting within a window size $\ell$ before calculating the velocity increment.  By doing so, the scales larger than $\ell$, 
i.e., $r\ge \ell$, are  expected to be removed or constrained.  The detrending process is equivalent to be a high-pass filter in physical domain.  Meanwhile the intermittency nature is retained.
We first validate the DSF method by using a synthesized fractional Brownian motion 
for mono-fractal processes and a lognormal process for multifractal random 
walk processes. The numerical  results show  comparable 
scaling exponents $\zeta(n)$ and singularity spectra $D(h)$ for the original SFs and DSFs. 
When applying the DSF to a turbulent velocity obtained from a high Reynolds number wind tunnel experiment with $Re_{\lambda}\simeq 720$, the 3rd-order DSF demonstrates  a clear inertial range with $\mathcal{B}_3(\ell)\simeq 4/5\epsilon \ell$ on the range  $10<\ell/\eta<1000$, corresponding to a wavenumber range $0.001<k\eta<0.1$. This inertial range  is consistent with the one predicted by  the Fourier power spectrum. 
The directly measured  scaling exponents $\zeta(n)$ (resp. singularity spectrum $D(h)$)   agree very well with a lognormal model with an intermittent parameter $\mu=0.33$.  Due to  large-scale 
 effects, the results provided by 
 the SFs  are biased. The method proposed here is general and 
 can be applied to different dynamics systems in which the concepts of 
multiscale and multifractal are relevant. 

\end{abstract}

\section{Introduction}

Multiscale dynamics is present in many phenomena, e.g., turbulence \cite{Frisch1995}, finance \cite{Schmitt2000,Muzy2001QF}, 
 geosciences \cite{Schmitt2009JMS,Lovejoy2012NPG}, etc, to quote a few. It has been found in many multiscale dynamics systems that the self-similarity is broken, in which the concept of multiscaling or multifractal is relevant \cite{Frisch1995}. This is characterized conventionally by using the 
 structure-functions (SFs), i.e., $S_n(\ell)=\langle \Delta u_{\ell}(x)^n \rangle \sim \ell ^{\zeta(n)}$, in which $\Delta u_{\ell}(x)= u(x+\ell)-u(x)$ is
 an increment with separation scale $\ell$.  Note that for the self-similarity process, e.g., fractional Brownian motion 
 (fBm), the measured $\zeta(n)$ is  linear with $n$. While for the multifractal 
 process, e.g., turbulent velocity, it is usually convex with $n$. Other methods are available to extract the scaling exponent. For example, wavelet based methodologies, (e.g., wavelet leaders, wavelet transform modulus maxima \cite{Lashermes2008EPJB,Muzy1993PRE,Lovejoy2012NPG}),  Hilbert-based method \cite{Huang2008EPL,Huang2011PRE}, or the scaling analysis of probability density  
 function of velocity increments \cite{Huang2011PoF}, to name a few. Each method has its 
 owner advantages and shortcomings. For example, the classical SFs is found to mix 
 information of the large- (resp. known as infrared effect) and small-scale (resp. known as ultraviolet effect) structures 
 \cite{Davidson2005PRL,Huang2010PRE,Blum2010PoF,Huang2011PRE,Huang2013PRE}. The corresponding scaling exponent 
 $\zeta(n)$ is thus biased when a large energetic structure is present \cite{Huang2011PRE}.

Previously the influence of the large-scale structure has been considered extensively by several authors \cite{PraskvoskyJFM1993,Sreenivasan1996PRL,Sreenivasan1998PTP,Huang2010PRE,Blum2010PoF,Blum2011NJP}.
For example, Praskvosky et al., \citep{PraskvoskyJFM1993} found strong correlations between the large scales and the velocity SFs at all length scales.   Sreenivasan \& Stolovitzky \citet{Sreenivasan1996PRL} observed that the inertial range of  the SFs  conditioned  on the large scale velocity show a strong dependence. Huang et al., \citet{Huang2010PRE} showed analytically
 that the influence of the large-scale structure could be as large as two decades down to the small scales. Blum et al., \citet{Blum2010PoF} studied experimentally the nonuniversal large-scale structure by considering both conditional Eulerian and Lagrangian SFs.  They found that both SFs  depend on the strength of  large-scale structures at all scales. In their study, 
 the large-scale structure velocity is defined as two-point average, i.e., $\sum u_z(\ell)=[u_z(x)+u_z(x+\ell)]/2$, in which $u_z$ is the vertical velocity in their experiment apparatus. Note that they conditioned  SFs on different intensity of $\sum u_z(\ell)$.  Later, Blum et al., \citet{Blum2011NJP} investigated systematically the large-scale structure conditioned SFs for various turbulent flows.  They confirmed  that in different turbulent flows the conditioned SFs depends strongly on large-scale structures at all scales.

In this paper, a detrended structure-function (DSF) method is proposed to 
extract scaling exponents $\zeta(n)$. This is accomplished by removing a $1$st-order 
polynomial within a window size $\ell$ before calculating the velocity increment. This  procedure  is designated as detrending analysis (DA). By doing so,  scales larger than $\ell$, 
i.e., $r\ge \ell$, are  expected to be removed or constrained.  Hence, the DA acts as 
a high-pass filter in physical domain. Meanwhile, the intermittency is still retained. A
velocity increment $\Delta u_{i,\ell}(x)$ is then defined within the window size $\ell$. A $n$th-order moment of $\Delta u_{i,\ell}(x)$ is  introduced as $n$th-order DSF. 
 The DSF is first validated by using a synthesized fractional Brownian motion (fBm) and a  lognormal process with an intermittent parameter $\mu=0.15$ respectively for mono-fractal and multifractal processes. It is found that DSFs provide  comparable scaling exponents $\zeta(n)$ and singularity spectra $D(h)$ with the ones provided by the original SFs.
When applying to a turbulent velocity with a Reynolds number $Re_{\lambda}=720$, the $3$rd-order DSF  shows a clear inertial range $10<\ell/\eta<1000$, which  is consistent with the one predicted by the Fourier power spectrum $E_u(k)$, e.g., $0.001<k\eta<0.1$.  Moreover, a compensated height of the $3$rd-order DSF is $0.80\pm0.05$. This value is consistent with the famous Kolmogorov four-fifth law. 
The directly measured
scaling exponents $\zeta(n)$ (resp. singularity spectrum $D(h)$)  agree very well with the lognormal model with an intermittent parameter $\mu=0.33$. 
Due to the large-scale effect, known as infrared effect, the SFs  are  biased.
  Note that  the scaling exponents are extracted directly  without resorting to the Extended-Self-Similarity (ESS) technique.
The method is general and could be applied to different types of data, in which the multiscale and multifractal concepts are relevant.

\section{Detrending Analysis and Detrended Structure-Function}

\subsection{Detrending Analysis}

We start here with  a scaling process $u(x)$, which has a power-law Fourier spectrum, i.e.,
\begin{equation}
E(k)=C k^{-\beta}
\end{equation}
in which $\beta$ is the scaling exponent of $E(k)$.  The Parseval's theorem states the following relation, i.e.,
\begin{equation}
\langle u(x)^2 \rangle_x = \int_{0}^{+\infty} E(k)\upd k\label{eq:Parseval1}
\end{equation}
in which $\langle \,\rangle$ is ensemble average, $E(k)$ is the Fourier power spectrum of $u(x)$ \cite{Percival1993}.
We first divide  the given $u(x)$ into  $m$ segments with a length $\ell$ each.
A $q$th-order detrending of the $i$th segment is defined as, i.e.,
\begin{equation}
u_{i,\ell}(x)=u_i(x)-P_{i,\ell}^q(x),\, (i-1)\ell\le x\le i\ell \label{eq:detrend}
\end{equation}
in which $P_{i,\ell}^q(x)$ is a $q$th-order polynomial fitting of the $u_i(x)$. 
We consider below only for the first-order detrending, i.e., $q=1$.
 To obtain a detrended signal, i.e., 
$u_{\ell}(x)=[u_{1,\ell}(x),u_{2,\ell}(x)\cdots u_{m,\ell}(x)]$, a linear trend  is removed within a window size $\ell$. 
Ideally, scales larger than $\ell$, i.e., $r>\ell$ are removed or constrained from the  original data $u(x)$. This implies that the DA procedure is  a high-pass filter in the physical domain.
The kinetic energy of $u_{\ell}(x)$ is related directly with its Fourier power spectrum, i.e.,
\begin{equation}
\mathcal{D}_2(\ell)=\langle u_{\ell}(x)^2 \rangle_x = \int_{0}^{+\infty} E_{\ell}(k)\upd k\simeq\int_{k_{\ell}}^{+\infty} E(k)\upd k\label{eq:Parseval2}
\end{equation}
in which $k_{\ell}=1/\ell$ and $E_{\ell}(k)$ is the Fourier power spectrum of $u_{\ell}(x)$.  
This illustrates again that the DA procedure acts a high-pass filter, in which the lower Fourier modes $k<k_{\ell}$ (resp. $r>\ell$) are expected to be removed or constrained. 
 For a scaling process, i.e., $E(k)\sim k^{-\beta}$, it leads a power-law behavior, i.e.,
\begin{equation}
\mathcal{D}_2(\ell)\sim k_{\ell}^{1-\beta}\sim \ell^{\beta-1}\label{eq:DA2}
\end{equation}
The physical meaning of $\mathcal{D}_2(\ell)$ is quite clear.  It represents 
a cumulative energy over the Fourier wavenumber band $[k_{\ell},+\infty]$ (resp. scale range $[0,\ell]$). We emphasize here again that the DA acts as a high-pass filter in physical domain and the intermittency nature of $u(x)$ is still retained.

\subsection{Detrended Structure-Function}

 The above mentioned detrending analysis can
remove/constrain the large-scale influence, known as infrared effect.  This could be utilized to redefine the SF to remove/constrain the large-scale structure effect as following. After the DA procedure, , the velocity increment can be defined  within  a window size $\ell$ as, i.e.,
\begin{equation}
\Delta u_{i,\ell}(x)= u_{i,\ell}(x+\ell/2)-u_{i,\ell}(x) \label{eq:DVI}
\end{equation} 
in which $i$ represents for the $i$th segment. 
We will show in the next subsection why we define an increment with a half width of the window size.
A $n$th-order DSF is then defined as, i.e.,
\begin{equation}
\mathcal{B}_n(\ell)=\langle \Delta u_{i,\ell}(x)^n\rangle_x\label{eq:DSF}
\end{equation}
For a scaling process, we expect a power-law behavior, i.e.,
\begin{equation}
\mathcal{B}_n(\ell)\sim \ell^{\zeta(n)}
\end{equation}
in which the scaling exponent $\zeta(n)$ is comparable with the one provided by the original SFs.

To access  negative orders of $n$ (resp. the right part of the singularity spectrum $D(h)$, see definition below),  the DSFs can be redefined as, i.e.,
\begin{equation}
\mathcal{B}_n(\ell)=\langle  X_{\ell}(i)^n\rangle\label{eq:RDSF}
\end{equation}
in which $X_{\ell}(i)=\langle \vert \Delta u_{i,\ell}(x)  \vert \rangle_{(i-1)\ell\le x\le i\ell}$ is local average for the $i$th segment. A power-law behavior is expected, i.e., $\mathcal{B}_n(\ell)\sim \ell^{\zeta(n)}$. It is found experimentally that when $q>0$, Eqs.\eqref{eq:DSF} and \eqref{eq:RDSF} provide the same scaling exponents $\zeta(n)$.  In the following we do not discriminate these two definitions for DSFs. 

\subsection{An Interpretation in Time-wavenumber Analysis Frame}

\begin{figure}[!htb]
\centering
 \includegraphics[width=0.65\linewidth,clip]{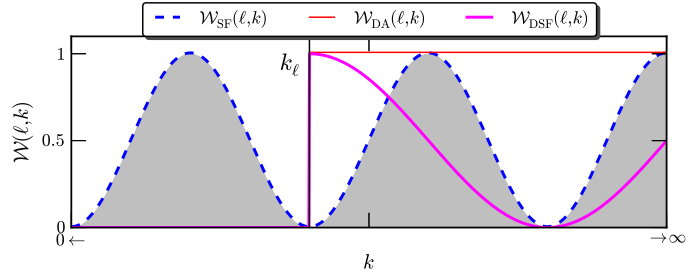}
  \caption{(Color online) An illustration of the weight function $\mathcal{W}(\ell,k)$ for different methods: structure-function $\mathcal{W}_{\textrm{SF}}=1-\cos(2\pi k \ell)$ (dashed line), first-order detrending analysis $\mathcal{W}_{\mathrm{DA}}$ (thin solid line), and the detrended structure-function $\mathcal{W}_{\mathrm{DSF}}$ (thick solid line).  The detrended scale $\ell$ is demonstrated by a vertical solid line with $k_{\ell}=1/\ell$. Ideally, scales larger than $\ell$, i.e., $r<\ell$ (resp. $k<k_{\ell}$) are expected to be removed after the detrending process.   }\label{fig:weight}
\end{figure}

To understand better the filter property of the detrending procedure and DSFs, we introduce here a weight function $\mathcal{W}(\ell,k)$, i.e.,
 \begin{equation}
M_2(\ell)=\int_{0}^{+\infty}
\mathcal{W}(\ell,k)E(k)\upd k\label{eq:weight}
 \end{equation} 
 in which $E(k)$ is the Fourier power spectrum of $u(x)$, and $M_2(\ell)$ is a second-order moment, which  could be one of
  $\mathcal{D}_2(\ell)$ or $\mathcal{B}_2(\ell)$, or  $S_2(\ell)$, respectively. 
  The weight function $\mathcal{W}(\ell,k)$ characterizes the contribution of the Fourier component to the corresponding second-order moment. 
  Note that an integral constant is neglected in the eq.\eqref{eq:weight}.
 For the second-order SFs,  one has the following weight function \cite{Frisch1995,Huang2010PRE}, i.e.,
 \begin{equation}
\mathcal{W}_{\mathrm{SF}}(\ell,k)=1-\cos\left(2\pi  k\ell \right)\label{eq:wSF}
 \end{equation} 
 For a scaling process, one usually has a fast decaying Fourier spectrum, i.e. $E(k)\sim k^{-\beta}$ with $\beta>0$. Hence, the contribution from small-scale (resp. high wavenumber Fourier mode) is decreasing.  The SFs  might be more influenced by the large-scale  part for large values of   $\beta$ \cite{Huang2010PRE,Huang2013PRE,Tan2014PoF}.
 For the detrended data, the corresponding weight function is ideally to be as the following, i.e.,
 \begin{equation}
\mathcal{W}_{\mathrm{DA}}(\ell,k)= \left\{
\begin{array}{lll}
&0, &\textrm{when $k\le k_{\ell}$}\\
&1, &\textrm{when $k>k_{\ell}$ }
\end{array}
\right.
\label{eq:wDA}
 \end{equation} 
 The DSFs (resp. the combination of the DA and SF) have a weight function, i.e.,
   \begin{equation}
\mathcal{W}_{\mathrm{DSF}}(\ell,k)=\left\{
\begin{array}{lll}
&0, &\textrm{when $k\le k_{\ell}$}\\
&1-\cos\left(\pi k \ell \right), &\textrm{when $k>k_{\ell}$ }
\end{array}
\right.
\label{eq:wDA}
 \end{equation} 
Comparing with the original SFs, the DSFs defined here can remove/constrain the large-scale effect.  Figure \ref{fig:weight} shows the corresponding $\mathcal{W}(\ell,k)$ for the SF, detrending analysis, 
and DSF, respectively.  The detrended scale $\ell$ is illustrated by a vertical line, 
i.e., $k_{\ell}=1/\ell$. 
We note here that with the definition of Eq.\eqref{eq:DVI},  $\mathcal{B}_2(\ell)$ 
provides a better compatible interpretation with the Fourier power spectrum $E(k)$ 
since we have $\mathcal{W}_{\mathrm{DSF}}(\ell,k_{\ell})=1$.   This is the main reason why we define the velocity increment with the half size of the window width $\ell$.
    
  We provide some comments on Eq.\eqref{eq:weight}.   The above argument is exactly valid for linear and stationary processes. In reality, the data are always nonlinear 
  and nonstationary for some reasons, see more discussion in Ref. 
  \cite{Huang1998EMD}.  Therefore, eq.\eqref{eq:weight} holds approximately for real data. 
  Another comment has to be 
  emphasized here for the detrending procedure. 
  Several approaches might  be applied 
  to remove the trend 
  \cite{Wu2007,Bashan2008}. 
However,  the trend might be linear or nonlinear. Therefore, different detrending approaches might provide different performances.  In the present study, we only consider  the $1$st-order polynomial detrending procedure, which is  efficient for many types of data.

\begin{figure}[!htb]
\centering
 \includegraphics[width=0.65\linewidth,clip]{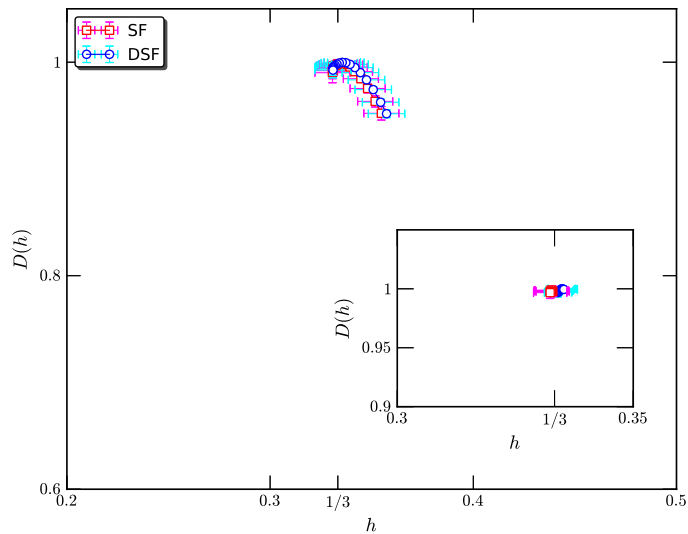}
  \caption{(Color online) Measured singularity spectrum $D(h)$ for fractional Brownian motion with $H=1/3$ on the range $-4\le n \le 4$. The inset shows the singularity spectrum $D(n)$ on the range $0\le n \le 4$. The errorbar is the standard deviation estimated from 100 realizations. Ideally, one should have $h=1/3$ and $D(1/3)=1$.  
Both methods provide the same  $h$ and $D(h)$ and statistical error. }\label{fig:fBm}
\end{figure}

\begin{figure}[!htb]
\centering
 \includegraphics[width=0.65\linewidth,clip]{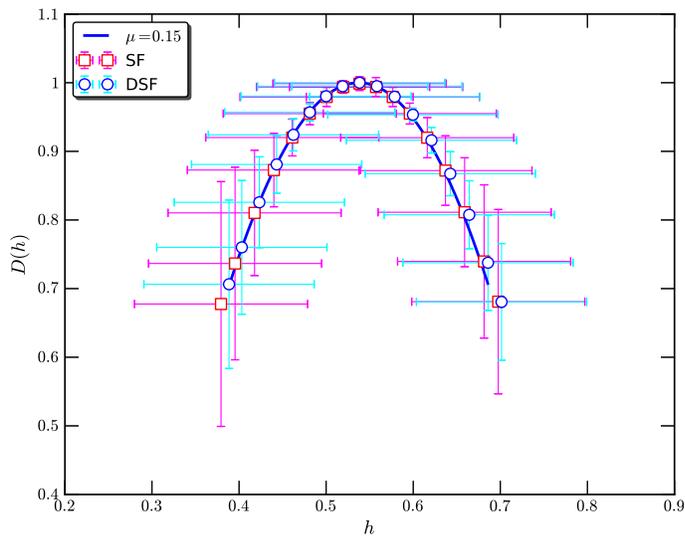}
  \caption{(Color online) Measured singularity spectrum $D(h)$ for the lognormal process with an intermittent parameter $\mu=0.15$. The errorbar is the standard deviation from the 100 realizations.  The theoretical singularity curve is illustrated by a solid line. Both estimators provide the same singularity spectra $D(h)$ and statistical error. }\label{fig:wfbm}
\end{figure}

\section{Numerical Validation}
\subsection{Fractional Brownian Motion}

We first consider here the fractional Brownian motion as a typical mono-scaling process. 
FBm  is a
Gaussian self-similar process with a normal distribution increment, which is
characterized by $H$,  namely  Hurst number $0<H<1$
\cite{Beran1994,Rogers1997,Doukhan2003,Gardiner2004}.  A Wood-Chan algorithm is used to synthesize  the fBm with a Hurst number 
$H=1/3$. We perform 100 realizations with a data length $10^5$ points each.  Power-law behavior is observed on a large-range of scales for $-4\le n\le 4$.
The corresponding singularity spectrum is, i.e., 
\begin{equation}
h=\zeta\rq{}(n),\,D(h)=\min_n\{ h n -\zeta(n)+1\}
\end{equation}
Ideally, one should have a single point of  singularity spectrum with $h=1/3$ and $D(1/3)=1$. However, in practice, the measured singularity spectrum $D(h)$  is always lying in a narrow band.
Figure \ref{fig:fBm} shows the measured singularity spectrum  $D(h)$ for SFs ($\square$) and DSFs ($\ocircle$) for $-4\le n\le 4$, in which the inset shows the singularity spectra $D(h)$ estimated on the range $0\le n\le 4$.  Visually,
both estimators provide the same  $D(h)$ and  the  same statistical error, which is defined as the standard deviation from different realizations.

\subsection{Multifractal Random Walk With a Lognormal Statistics}

We now  consider a multifractal random walk with a lognormal statistics \cite{Bacry2001,Muzy2002,Schmitt2003}.
A multiplicative discrete  cascade process with a lognormal statistics is performed to simulate a multifractal  measure $\epsilon(x)$. The larger scale corresponds to a unique cell of size $L=\ell_0 \lambda_1^N$, where $\ell_0$ is the largest scale considered and $\lambda_1>1$ is   a  dimensional scale ratio. In practice for a discrete model, this ratio is often taken as $\lambda_1=2$ \cite{Schmitt2003,Huang2011PRE}. The next scale involved 
corresponds to $\lambda_1$ cells, each of size $L/\lambda_1=\ell_0 \lambda_1^{N-1}$. This is iterated and at step $p$ ($1\le p \le N$)  $\lambda_1^p$ cells are retrieved.
Finally, at each point the multifractal measure $\epsilon(x)$  is  as the product of 
$n$ cascade random variables, i.e.,
\begin{equation}
  \epsilon(x)=\prod_{m=1}^N W_m(x)
\end{equation}
where $W_{m}(x)$ is the random variable corresponding to position $x$ and level $m$ in the cascade \cite{Schmitt2003}.
Following the multifractal random
walk idea \cite{Bacry2001,Muzy2002},  a nonstationary multifractal time series can be synthesized  as,  i.e.,
\begin{equation}
 u(x)=\int_{0}^x {\epsilon(x')^{1/2}} \upd B(x')\label{eq:multitime}
\end{equation}
where $B(x)$ is Brownian motion. Taking a  lognormal statistic for $\epsilon$,
the scaling exponent $\zeta(n)$ for the SFs, i.e., $\langle \Delta u_{\ell}(x)^n\rangle\sim
\ell^{\zeta(n)}$, is written as, 
\begin{equation}
\zeta(n)=\frac{n}{2}-\frac{\mu}{2}(\frac{n^2}{4}-\frac{n}{2})\label{eq:lognormal}
\end{equation}
where $\mu$ is the  intermittency parameter ($0\le\mu\le1$) characterizing
the lognormal multifractal cascade \cite{Schmitt2003}.

Synthetic multifractal time series are generated following 
Eq.\eqref{eq:multitime}. 
An intermittent parameter $\mu=0.15$ is chosen for $m=17$ levels each, corresponding to a data length $131072$ points each. A total of
100 realizations are performed. The statistical error is then measured as the standard 
deviation from these realizations.  Figure \ref{fig:wfbm} shows the 
corresponding measured singularity spectra $D(h)$, in which the theoretical value is 
illustrated by a solid line. Graphically, the theoretical  singularity spectra $D(h)$ are recovered by both estimators. Statistical error are again found to be the same for both estimators.

We would like to provide some comments on the performance of these 
two estimators.  For the synthesized processes, they  have the same 
performance since there is no intrinsic 
structure  in these synthesized data. But for the real data, as we mentioned 
above, they possess nonstationary and nonlinear structures 
\cite{Huang1998EMD}.  Therefore, as shown in below, they might have different performance.

\section{Application to Turbulent Velocity}

\begin{figure}[!htb]
\centering
 \includegraphics[width=0.65\linewidth,clip]{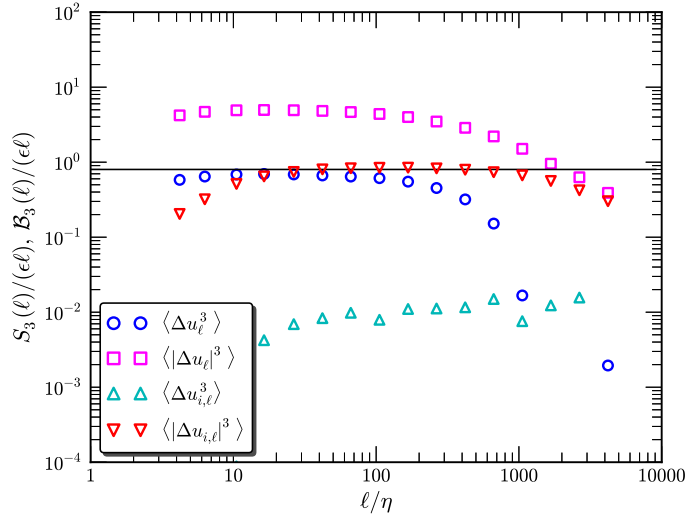}
  \caption{(Color online) Measured compensated 3rd-order moments $S_3(\ell)/(\epsilon\ell)$ and $\mathcal{B}_3(\ell)/(\epsilon\ell)$ from experimental homogeneous and nearly isotropic turbulent flow. They are respectively 3rd-order SFs with ($\square$) and without ($\ocircle$) absolute value, and 3rd-order DSFs with ($\triangledown$) and without ($\triangle$) absolute value.  The horizontal solid line indicates the Kolmogorov\rq{}s four-fifth law. An observed plateau for $\mathcal{B}_3(\ell)/{\epsilon \ell}$ indicates an inertial range  on the range $10<\ell/\eta<1000$, corresponding to a wavenumber range $0.001<k\eta<0.1$. Roughly speaking, a plateau for $S_3(\ell)/{\epsilon\ell}$ indicates an inertial range on the range $10<\ell/\eta<100$.
  The height of the inertial range are respectively $0.67\pm0.02$ ($\ocircle$), $4.84\pm0.14$ ($\square$),  $0.0098\pm0.0024$ ($\triangle$) and $0.80\pm0.05$ ($\triangledown$), in which the statistical error is the standard deviation obtained from the inertial range. Note that the inertial range  are $10<\ell/\eta<100$ for the SFs and $10<\ell/\eta<1000$ for the DSFs. The corresponding scaling exponents $\zeta(3)$ are $0.95\pm0.02$, $0.84\pm 0.03$, $1.15\pm0.07$ and $0.99\pm0.03$. The statistical error is the  95\% fitting confidence  on the inertial range.
   }\label{fig:third}
\end{figure}

\begin{figure}[!htb]
\centering
 \includegraphics[width=0.65\linewidth,clip]{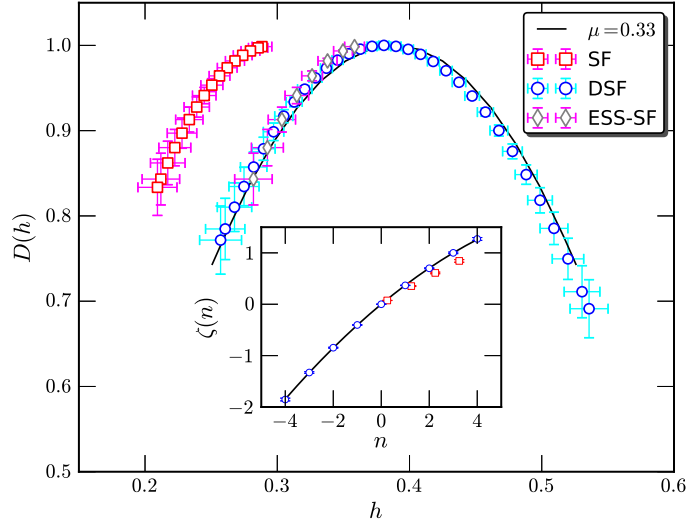}
  \caption{(Color online) Measured singularity spectrum $D(h)$. The errorbar is the standard deviation from 120 realizations. The inset shows the corresponding scaling exponents $\zeta(n)$. For comparison, the lognormal model with an intermittent parameter $\mu=0.33$ is illustrated by a solid line. 
   }\label{fig:singularity}
\end{figure}

We consider here a velocity database obtained from a high Reynolds number wind tunnel experiment in the Johns-Hopkins university
with 
Reynolds number $Re_{\lambda}=720$.  An probe array with four X-type hot wire anemometry is used to record 
the velocity with a  sampling wavenumber of $40$\,kHz at streamwise direction $x/M=20$, in which 
$M$ is the size of the active grid.   
These probes are placed in the middle height and along the center line of the wind tunnel  to record the turbulent velocity simultaneously for a duration of 30 second. The measurement is then repeated for 30 times. Finally, we have $30\times 4\times 30\times (4\times 10^4)$ data points (number of measurements $\times$ number of probes $\times$ duration  time $\times$ sampling wavenumber). Therefore, there are 120 realizations  (number of measurements $\times$ number of probes).
The Fourier power spectrum 
$E_u(k)$ of the longitudinal velocity reveals a nearly two decades inertial range on the wavenumber range $0.001<k\eta<0.1$ with a scaling exponent $\beta\simeq 1.65\pm0.02$, see Ref. \cite{kang2003}. This corresponds to  time scales $10<\ell/\eta<1000$.  Here $\eta$ is the Kolmogorov scale.
Note 
that we  convert our results into spatial space by applying 
the Taylor\rq{}s frozen hypothesis \cite{Frisch1995}. More detail about this database can be found in Ref.\,\cite{kang2003}.

To determine the inertial range in real space, we plot the measured compensated 
3rd-order moments in 
Fig.\ref{fig:third} for the SFs ($S_3(\ell)/(\epsilon\ell)$ with 
($\square$) and without ($\ocircle$) absolute value), DSFs 
($\mathcal{B}_3(\ell)/(\epsilon\ell)$ with ($\triangledown$) and 
without ($\triangle$) absolute value), respectively. 
A  horizontal solid line indicates the Kolmogorov\rq{}s four-fifth law. A
plateau is observed for $\mathcal{B}_3(\ell)/(\epsilon\ell)$ on the range 
$10<\ell/\eta<1000$, which agrees very well with the inertial range 
predicted by $E_u(k)$, i.e., on the range $0.001<k\eta<0.1$. The 
corresponding height and scaling exponent are $0.80\pm0.05$ with 
absolute value  (resp. $0.0098\pm0.0024$ without absolute value) and 
$\zeta(3)=0.99\pm0.03$ (resp. $\zeta(3)=1.15\pm0.07$), respectively. The statistical error 
is the standard deviation obtained from the range $10<\ell/\eta<1000$.  
Note that the Kolmogorov\rq{}s four-fifth law indicates a linear relation 
$\langle\Delta u_{\ell}^3\rangle=-4/5\epsilon\ell$. It is interesting to 
note that, despite of the sign, we have $\langle \vert \Delta u_{i,\ell}\vert 
^3\rangle=4/5\epsilon\ell$ on nearly two-decade scales. For comparison, 
the 3rd-order SFs are also shown. Roughly speaking, a plateau is observed 
on the range $10<\ell/\eta<100$. This inertial range is shorter than the 
one predicted by the Fourier analysis or DSFs, which is now understood as 
the large-scale influence. The corresponding height and scaling exponent 
are $0.67\pm0.02$ without absolute value (resp. $4.84\pm0.14$ with absolute value) and 
$0.95\pm0.02$ (resp. $0.84\pm0.03$). Therefore, the DSFs provide a better  indicator of the inertial range since it removes/constrains the large-scale influence.
We therefore estimate the scaling exponents 
for the $\mathcal{B}_n(\ell)$ on the range $10<\ell/\eta<1000$ for $-4\le n \le 4$ 
directly without resorting to the Extended Self-Similarity technique \cite{Benzi1993PRE,Benzi1993EPL}. 
For the SFs, we  calculate the scaling exponents $\zeta(n)$ on the range $10<\ell/\eta<100$ for $0\le n\le 4$ directly.

 Figure \ref{fig:singularity} shows the measured singularity spectra $D(h)$ for $-4\le n\le 4$, in which the errorbar is a standard deviation from 120 realizations. The inset shows the corresponding scaling exponents $\zeta(n)$. For comparison, the lognormal model $\zeta(n)=n/3-\mu/18\left(n^2-3n \right)$ with an intermittent parameter $\mu=0.33$  is shown as a solid line.  
  Visually, the DSFs curve fully recovers the lognormal curve not only on the left part (resp. $n\ge 0$) but also on the right part (resp. $n\le 0$).
   Due to the large-scale contamination, the SFs underestimates the scaling exponents $\zeta(n)$ when $n\ge 0$ \cite{Davidson2005PRL,Huang2010PRE}. This 
   leads an overestimation of the left part of singularity spectrum $D(h)$ (see 
   $\square$ in Fig.\ref{fig:singularity}). However, if one resorts the ESS algorithm when measuring the SF scaling exponent $\zeta(n)$, the 
   corresponding singularity spectrum $D(h)$ is then horizontal shifted to the 
   theoretical curve. This has been interpreted as that the ESS technique suppresses
   the finite Reynolds number effect. We show here that if one removes/constrains  the effect of
 large-scale  motions, one can retrieve the scaling exponent $\zeta(n)$ (resp. singularity spectrum $D(h)$) without resorting the ESS technique.   Or in other words, the finite Reynolds number effect manifests at  large-scale motions, which  is usually anisotropic too.

\section{Conclusion}
In this paper, we introduce a detrended structure-function analysis to 
remove/constrain the influence of large-scale motions, known as the infrared 
effect.   In the first step  of our proposal, the $1$st-order polynomial 
trend is removed within a window size $\ell$. By doing so, the scales 
larger than $\ell$, i.e., $r\ge\ell$, are expected to be removed/constrained. In the
second step, a velocity increment is
defined with a   half of the window size. The DSF proposal is validated by the synthesized fractional 
Brownian motion for the mono-fractal process and a lognormal  random walk 
for the multifractal process.  The numerical test shows that both  SFs and DSFs
estimators 
provide a comparable performance for  synthesized processes without 
intrinsic structures.  
 
 When applying to the turbulent velocity obtained from a high Reynolds
 number wind tunnel experiment, the 3rd-order DSFs show a clearly 
 inertial range  on the range $10<\ell/\eta<1000$ with a linear relation $\mathcal{B}_3(\ell)\simeq 4/5\epsilon\ell$. The inertial range provided by DSFs is consistent with the one
 predicted by the Fourier power spectrum. 
Note that, despite of the sign, the Kolmogorov\rq{}s four-fifth law is  retrieved for the 3rd-order DSFs.
 The corresponding 3rd-order 
 SFs  are biased by the large-scale structures, known as the infrared effect. It shows a shorter
 inertial range and  underestimate  the 3rd-order scaling exponent $\zeta(3)$.  The
 scaling exponents $\zeta(n)$ are then estimated directly without resorting to the ESS 
 technique.  The corresponding singularity spectrum $D(h)$ provided 
 by the DSFs  fully recovers the lognormal model with an intermittent parameter $\mu=0.33$ on the range 
 $-4\le n \le 4$. However, the classical SFs  overestimate the  left part 
 singularity spectrum $D(h)$ (resp. underestimate the corresponding  scaling 
 exponents $\zeta(n)$) on the range $0\le n\le4$. This has been interpreted as finite 
 Reynolds number effect and can be corrected by using the ESS technique.  
 Here, to our knowledge, we show for the first time that if one 
 removes/constrains the influence of the large-scale structures, one can
 recover the lognormal model without resorting to the ESS technique.

The method we proposed here is general and applicable to other complex dynamical systems, in which the multiscale statistics are relevant. It  should be also applied systematically to more turbulent velocity databases with different Reynolds numbers to  see whether the finite Reynolds number effect manifests on  large-scale motions as well as we show for high Reynolds number turbulent flows.

\section*{Acknowledgements}
 This work is sponsored by the National Natural Science Foundation of China under Grant (Nos.  11072139, 11032007,11161160554, 11272196, 11202122 and 11332006) ,  \lq{}Pu Jiang\rq{} project of Shanghai (No. 12PJ1403500), Innovative program of Shanghai Municipal Education Commission
(No. 11ZZ87)
  and  the 
 Shanghai Program for Innovative Research Team in Universities. Y.H. thanks Prof. F.G. Schmitt for useful comments and suggestions.
We thank Prof. Meneveau for sharing his experimental
velocity database, which is available for download at
C. Meneveau's web page: {http://www.me.jhu.edu/meneveau/datasets.html}.
We  thank the  two anonymous referees for their useful comments and suggestions.

\bibliographystyle{tJOT}

\begin{thebibliography}{33}
\providecommand{\natexlab}[1]{#1}

\bibitem[1]{Frisch1995}
U. Frisch {\itshape {Turbulence: the legacy of AN Kolmogorov}},    Cambridge
  University Press, 1995.

\bibitem[2]{Schmitt2000}
F. Schmitt, D. Schertzer, and S. Lovejoy, {\itshape {Multifractal fluctuations
  in finance}}, Int. J. Theor. Appl. Fin 3 (2000), pp. 361--364.

\bibitem[3]{Muzy2001QF}
J. Muzy, D. Sornette, J. Delour, and A. Arneodo, {\itshape Multifractal returns
  and hierarchical portfolio theory}, Quant. Finance 1 (2001), pp. 131--148.

\bibitem[4]{Schmitt2009JMS}
F. Schmitt, Y. Huang, Z. Lu, Y. Liu, and N. Fernandez, {\itshape Analysis of
  velocity fluctuations and their intermittency properties in the surf zone
  using empirical mode decomposition}, J. Mar. Sys. 77 (2009), pp. 473--481.

\bibitem[5]{Lovejoy2012NPG}
S. Lovejoy, and D. Schertzer, {\itshape Haar wavelets, fluctuations and
  structure functions: convenient choices for geophysics}, Nonlinear Proc.
   Geoph. 19 (2012), pp. 513--527.

\bibitem[6]{Lashermes2008EPJB}
B. Lashermes, S. Roux, P. Abry, and S. Jaffard, {\itshape {Comprehensive
  multifractal analysis of turbulent velocity using the wavelet leaders}}, Eur.
  Phys. J. B 61 (2008), pp. 201--215.

\bibitem[7]{Muzy1993PRE}
J. Muzy, E. Bacry, and A. Arneodo, {\itshape {Multifractal formalism for
  fractal signals: The structure-function approach versus the wavelet-transform
  modulus-maxima method}}, Phys. Rev. E 47 (1993), pp. 875--884.

\bibitem[8]{Huang2008EPL}
Y. Huang, F. Schmitt, Z. Lu, and Y. Liu, {\itshape An amplitude-frequency study
  of turbulent scaling intermittency using Hilbert spectral analysis},
  Europhys. Lett. 84 (2008), p. 40010.

\bibitem[9]{Huang2011PRE}
Y. Huang, F.G. Schmitt, J.P. Hermand, Y. Gagne, Z. Lu, and Y. Liu, {\itshape
  Arbitrary-order Hilbert spectral analysis for time series possessing scaling
  statistics: comparison study with detrended fluctuation analysis and wavelet
  leaders}, Phys. Rev. E 84 (2011), p. 016208.

\bibitem[10]{Huang2011PoF}
Y. Huang, F. Schmitt, Q. Zhou, X. Qiu, X. Shang, Z. Lu, and Y. Liu, {\itshape
  {Scaling of maximum probability density functions of velocity and temperature
  increments in turbulent systems}}, Phys. Fluids 23 (2011), p. 125101.

\bibitem[11]{Davidson2005PRL}
P.A. Davidson, and B.R. Pearson, {\itshape Identifying turbulent energy
  distribution in real, rather than Fourier, space}, Phys. Rev. Lett. 95
  (2005), p. 214501.

\bibitem[12]{Huang2010PRE}
Y. Huang, F. Schmitt, Z. Lu, P. Fougairolles, Y. Gagne, and Y. Liu, {\itshape
  {Second-order structure function in fully developed turbulence}}, Phys. Rev.
  E 82 (2010), p. 026319.

\bibitem[13]{Blum2010PoF}
D.B. Blum, S.B. Kunwar, J. Johnson, and G.A. Voth, {\itshape Effects of
  nonuniversal large scales on conditional structure functions in turbulence},
  Phys. Fluids 22 (2010), p. 015107.

\bibitem[14]{Huang2013PRE}
Y. Huang, L. Biferale, E. Calzavarini, C. Sun, and F. Toschi, {\itshape
  Lagrangian single particle turbulent statistics through the Hilbert-Huang
  Transforms}, Phys. Rev. E 87 (2013), p. 041003(R).

\bibitem[15]{PraskvoskyJFM1993}
A.A. Praskovsky, E.B. Gledzer, M.Y. Karyakin, and Y. Zhou, {\itshape The
  sweeping decorrelation hypothesis and energy-inertial scale interaction in
  high Reynolds number flows}, J. Fluid Mech. 248 (1993), p. 493.

\bibitem[16]{Sreenivasan1996PRL}
K.R. Sreenivasan, and G. Stolovitzky, {\itshape Statistical dependence of
  inertial range properties on large scales in a high-Reynolds-number shear
  flow}, Phys. Rev. Lett. 77 (1996), p. 2218.

\bibitem[17]{Sreenivasan1998PTP}
K.R. Sreenivasan, and B. Dhruva, {\itshape {Is there scaling in high-Reynolds
  number turbulence?}}, Prog. Theor. Phys. 130 (1998), pp. 103--120.

\bibitem[18]{Blum2011NJP}
D.B. Blum, G.P. Bewley, E. Bodenschatz, M. Gibert, A. Gylfason, L. Mydlarski,
  G.A. Voth, H. Xu, and P. Yeung, {\itshape Signatures of non-universal large
  scales in conditional structure functions from various turbulent flows}, New
  J. Phys. 13 (2011), p. 113020.

\bibitem[19]{Percival1993}
D. Percival, and A. Walden {\itshape {Spectral Analysis for Physical
  Applications: Multitaper and Conventional Univariate Techniques}},
  Cambridge University Press, 1993.

\bibitem[20]{Tan2014PoF}
H. Tan, Y. Huang, and J.P. Meng, {\itshape Hilbert Statistics of Vorticity
  Scaling in Two-Dimensional Turbulence}, Phys. Fluids 26 (2014), p. 015106.

\bibitem[21]{Huang1998EMD}
N. Huang, Z. Shen, S. Long, M. Wu, H. Shih, Q. Zheng, N. Yen, C. Tung, and H.
  Liu, {\itshape The empirical mode decomposition and the Hilbert spectrum for
  nonlinear and non-stationary time series analysis}, Proc. R. Soc. London,
  Ser. A 454 (1998), pp. 903--995.

\bibitem[22]{Wu2007}
Z. Wu, N.E. Huang, S.R. Long, and C. Peng, {\itshape {On the trend, detrending,
  and variability of nonlinear and nonstationary time series}}, PNAS 104
  (2007), p. 14889.

\bibitem[23]{Bashan2008}
A. Bashan, R. Bartsch, J. Kantelhardt, and S. Havlin, {\itshape {Comparison of
  detrending methods for fluctuation analysis}}, Physica A 387 (2008), pp.
  5080--5090.

\bibitem[24]{Beran1994}
J. Beran {\itshape {Statistics for long-memory processes}},    CRC Press, 1994.

\bibitem[25]{Rogers1997}
L. Rogers, {\itshape {Arbitrage with Fractional Brownian Motion}}, Math.
  Finance 7 (1997), pp. 95--105.

\bibitem[26]{Doukhan2003}
P. Doukhan, M. Taqqu, and G. Oppenheim {\itshape {Theory and Applications of
  Long-Range Dependence}},    Birkhauser, 2003.

\bibitem[27]{Gardiner2004}
C.W. Gardiner {\itshape Handbook of Stochastic Methods},    Springer, Berlin,
  third edition, 2004.

\bibitem[28]{Bacry2001}
E. Bacry, J. Delour, and J. Muzy, {\itshape {Multifractal random walk}}, Phys.
  Rev. E 64 (2001).

\bibitem[29]{Muzy2002}
J. Muzy, and E. Bacry, {\itshape {Multifractal stationary random measures and
  multifractal random walks with log infinitely divisible scaling laws}}, Phys.
  Rev. E 66 (2002), p. 056121.

\bibitem[30]{Schmitt2003}
F. Schmitt, {\itshape {A causal multifractal stochastic equation and its
  statistical properties}}, Eur. Phys. J. B 34 (2003), pp.
  85--98.

\bibitem[31]{kang2003}
H. Kang, S. Chester, and C. Meneveau, {\itshape {Decaying turbulence in an
  active-grid-generated flow and comparisons with large-eddy simulation}}, J.
  Fluid Mech. 480 (2003), pp. 129--160.

\bibitem[32]{Benzi1993PRE}
R. Benzi, S. Ciliberto, R. Tripiccione, C. Baudet, F. Massaioli, and S. Succi,
  {\itshape {Extended self-similarity in turbulent flows}}, Phys. Rev. E 48
  (1993), pp. 29--32.

\bibitem[33]{Benzi1993EPL}
R. Benzi, S. Ciliberto, C. Baudet, G. Chavarria, and R. Tripiccione, {\itshape
  {Extended self-similarity in the dissipation range of fully developed
  turbulence}}, Europhys. Lett 24 (1993), pp. 275--279.

\end{thebibliography}

\end{document}